\documentclass[%
 reprint,
superscriptaddress,
showpacs,
 amsmath,amssymb,
 aps,
 pra,
 floats,
]{revtex4-1}

\usepackage[latin1]{inputenc}
\usepackage{graphicx}
\usepackage{dcolumn}
\usepackage{bm}
\usepackage{units}
\usepackage{color}

\newcommand{\ket}[1]{ \left|#1\right\rangle}
\renewcommand{\vec}[1]{\bm{#1}}
\newcommand{\tens}[1]{\mbox{\textbf{\textsf{\textit{#1}}}}}

\newcommand{\sprod}{\cdot}
\newcommand{\tprod}{}

\newcommand{\trans}{{\operatorname{T}}}
\newcommand{\dif}{\mathrm{d}}
\newcommand{\mi}{\mathrm{i}}
\newcommand{\me}{\mathrm{e}}

\newcommand{\bra}[1]{\left\langle{#1}\right|}

\begin{document}

\preprint{APS/123-QED}

\title{Directional spontaneous emission and lateral Casimir-Polder
force on an atom close to a nanofiber}

\author{Stefan Scheel}
\affiliation{Institut f\"ur Physik, Universit\"at Rostock, Universit\"atsplatz
3, D-18055 Rostock, Germany}
\author{Stefan Yoshi Buhmann}
\affiliation{Physikalisches Institut, Albert-Ludwigs-Universit\"at Freiburg,
Hermann-Herder-Str. 3, D-79104 Freiburg, Germany}
\affiliation{Freiburg Institute for Advanced Studies,
Albert-Ludwigs-Universit\"at Freiburg, Albertstr. 19, D-79104 Freiburg, Germany}
\author{Christoph Clausen}
\author{Philipp Schneeweiss}
\affiliation{%
 Vienna Center for Quantum Science and Technology,
 TU Wien -- Atominstitut, Stadionallee 2, A-1020 Vienna, Austria
}%

\date{\today}

\begin{abstract}
We study the spontaneous emission of an excited atom close to an
optical nanofiber and the resulting scattering forces. For a suitably
chosen orientation of the atomic dipole, the spontaneous emission
pattern becomes asymmetric and a resonant Casimir--Polder force parallel
to the fiber axis arises.
For a simple model case, we show that the such a lateral force is
due to the interaction of the circularly oscillating atomic dipole
moment with its image inside the material.
With the Casimir--Polder energy being constant in the lateral
direction, the predicted lateral force does not derive from a
potential in the usual way. Our results have implications for optical
force measurements on a substrate as well as for laser cooling of atoms in
nanophotonic traps.
\end{abstract}

\maketitle

Electromagnetic fields close to a surface and their interaction with
particles are of fundamental interest. Research in this area
covers, e.g., dispersive interactions such as Casimir and
Casimir--Polder (CP) forces~\cite{Buhmann12}, as well as forces that
arise from the scattering of external light fields. Recently, optical
forces which act perpendicularly to the propagation direction of an
excitation light field attracted increasing interest~\cite{Wang14}. In
particular, it has been predicted that evanescent fields exert lateral
forces and torques on Mie particles~\cite{Bliokh14}. Even stronger
lateral forces are expected for chiral particles in evanescent
fields~\cite{ArXiv_Hayat14}.

Recent experimental research on emitters close to surfaces
have revealed  that
suitably excited particles can be used to realize strongly directional
excitation of guided
modes~\cite{Lin13,Rodriguez-Fortuno13,Luxmoore13a,
Neugebauer14,Mitsch14b,Petersen14,Feber15,ArXiv_Soellner14}.
For example, when a gold nanoparticle on the surface of an optical nanofiber
scatters the light of an external circularly polarized laser beam, the coupling
into counter-propagating guided modes of the nanofiber can exceed a ratio of
40:1~\cite{Petersen14}. Such an asymmetric scattering is independent of the
excitation process and is only governed by the polarization of the emitted
light. The conservation of total momentum in the system in conjunction with the
asymmetric emission suggests the existence of a force on the scatterer that is
parallel to the waveguide axis. However, the coupling to radiative modes has to
be taken into account as well~\cite{Xi13}. The search for such a scattering
force is within reach of current cold atom experiments, in which asymmetric
excitation of guided modes has already been observed~\cite{Mitsch14b}.

Lateral forces have also been discussed within the context of dispersion
interactions where they arise even in the absence of external fields.
Lateral CP forces are typically achieved by breaking the translational
invariance of the surface via periodic corrugations
\cite{Dalvit08,Doebrich08,Messina09,Contreras10} or
disorder \cite{Moreno10}.
Moreover, lateral Casimir forces between two periodically structured
surfaces have been proposed~\cite{Rodrigues06,Lambrecht08,Chiu09} and
measured~\cite{Chen02}, and have been put forward as a means to realize
contactless force transmission~\cite{Ashourvan07}.

\begin{figure}
\includegraphics[width=\columnwidth]{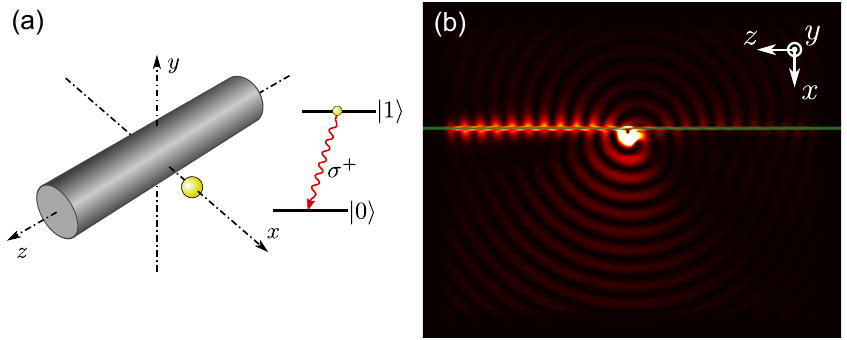}
\caption{(a) Sketch of the situation under consideration. An atom
  (yellow sphere) is located at \mbox{$\vec{r}_{\!A}=(x_{\!A},0,0)$},
  in close proximity to a nanofiber, modeled as a dielectric cylinder
  in free space. Calculations are performed for the cesium $D_2$
  line optical transition at a wavelength of $\lambda_{10}=852$~nm.
  The atom is initially in the excited state $|1\rangle$, from where
  it can only
  decay by emitting a $\sigma^+$-polarized photon. The cylinder
  consists of silica, has a radius \mbox{$R=250\,\mathrm{nm}$} and a
  refractive index $n=1.45$. The $y$-axis is taken as quantization
  axis. (b) Schematic emission pattern of a circular dipole antenna
  close to a waveguide obtained from a finite-difference time-domain
  simulation. Shown is a heat-map plot of the electric field energy
  density. A part of the emitted light is coupled into the waveguide
   (green horizontal line, width of 100~nm) and shows a strongly directional
  behavior, as expected~\cite{Mitsch14b}. The emission into radiation modes is
  asymmetric as well. For simplicity, the simulation is performed in
  two dimensions. For illustrative purposes only, we assumed a
  refractive index of the waveguide of $n=3>1.45$, which yields a
  larger asymmetry in the emission.  }
\label{fig:schematic}
\end{figure}

Here, we propose to exploit the directional spontaneous emission by an
atom near a nanofiber to realize a translationally invariant lateral CP force.
The envisioned physical situation is illustrated in Fig.~\ref{fig:schematic}(a)
and an illustration of an emission pattern is shown in
Fig.~\ref{fig:schematic}(b). A cesium atom is located at a position
\mbox{$\vec{r}_{\!A}=x_{\!A} \hat{\vec{x}}$} at a distance \mbox{$d_{\!A}=x_{\!A}-R$}
from the surface of a fused-silica fiber of radius \mbox{$R=250\,\mathrm{nm}$}.
The atom is initially prepared in an excited state \mbox{$|1\rangle
\equiv|6{}^2\mathrm{P}_{3/2},F'\!=\!5,M'_F\!=\!5\rangle$ }.
Here, the $y$ coordinate axis is chosen as the quantization axis. The
only available decay channel is to the ground state
\mbox{$|0\rangle\equiv|6{}^2\mathrm{S}_{1/2},%
F\!=\!4,M_F\!=\!4\rangle$}, such that the decay of the atom leads to
the emission of a $\sigma^+$-polarized photon. The transition has a free-space
wavelength of \mbox{$\lambda_{10}=852\,\mathrm{nm}$}, and corresponding
wavenumber $k_{10}=2\pi/\lambda_{10}$ and frequency $\omega_{10} = ck_{10}$.
The transition dipole matrix element is $\vec{d}_{10} = 1.9\times
10^{-29}\,\mathrm{Cm} \times (\mi \hat{\vec{x}} + \hat{\vec{z}})$~\cite{Steck10}.

We use two methods to calculate the emission rates and Casimir--Polder
force. The first method allows an intuitive interpretation and is
based on solving the Schrödinger equation for the atom-field
interaction for a suitable mode decomposition of the electric field
operator~\cite{LeKien05c}. The second method uses the more general
Green's tensor formalism where, for example, absorption of the
nanofiber is easily included.

The total spontaneous decay rate of an atom can be expressed in terms
of the Green's tensor $\tens{G}(\vec{r}, \vec{r}', \omega)$ as
\cite{Buhmann12}
\begin{equation}
  \label{eq:decay_greens}
  \Gamma(\vec{r}_{\!A}) = (2\mu_0/\hbar) \omega_{10}^2 \vec{d}_{10}\sprod
  \operatorname{Im}\tens{G}(\vec{r}_{\!A},\vec{r}_{\!A},\omega_{10})
  \sprod\vec{d}_{01}.
\end{equation}
A lateral force on the atom may result from an unbalanced spontaneous
emission rate into the $+z$ and $-z$ half spaces. To obtain
information about the directionality of the spontaneous emission, we
decompose the electric field into a set of orthonormal guided and
radiation modes of the nanofiber~\cite{LeKien05c}. By solving the
Schrödinger equation for the spontaneous emission problem~\cite{Supplement}, we obtain a total emission rate that is the sum of
partial decay rates $\gamma_{fp}^\text{(G)}$ into guided modes, and
$\gamma_{k_z m p}^\text{(R)}$ into radiation modes. Here, the index
$p$ labels the polarization, $f = \pm 1$ indicates the propagation
direction of the guided modes, $k_z$ is the projection of the wave
vector of the radiation mode onto the fiber axis, and $m \in
\mathbb{Z}$ is the mode order.
In general, the partial decay rates depend on $\vec{r}_{\!A}$. We
calculate the overall emission rates into the positive and negative
half spaces as
\begin{equation}
  \label{eq:RadDecay}
  \begin{split}
    \gamma_{\pm}^\text{(G)}(\vec{r}_{\!A}) &= \sum_p
    \gamma^\text{(G)}_{f=\pm 1, p}(\vec{r}_{\!A}),\\
    \gamma_{\pm}^\text{(R)}(\vec{r}_{\!A}) &= \pm \sum_{m,p}
    \int_0^{\pm k_{10}} \!\! \dif k_z\, \gamma^\text{(R)}_{k_z m p}(\vec{r}_{\!A}),
  \end{split}
\end{equation}
where the subscript indicates into which half space the emission is directed.

\begin{figure}
\includegraphics[width=\columnwidth]{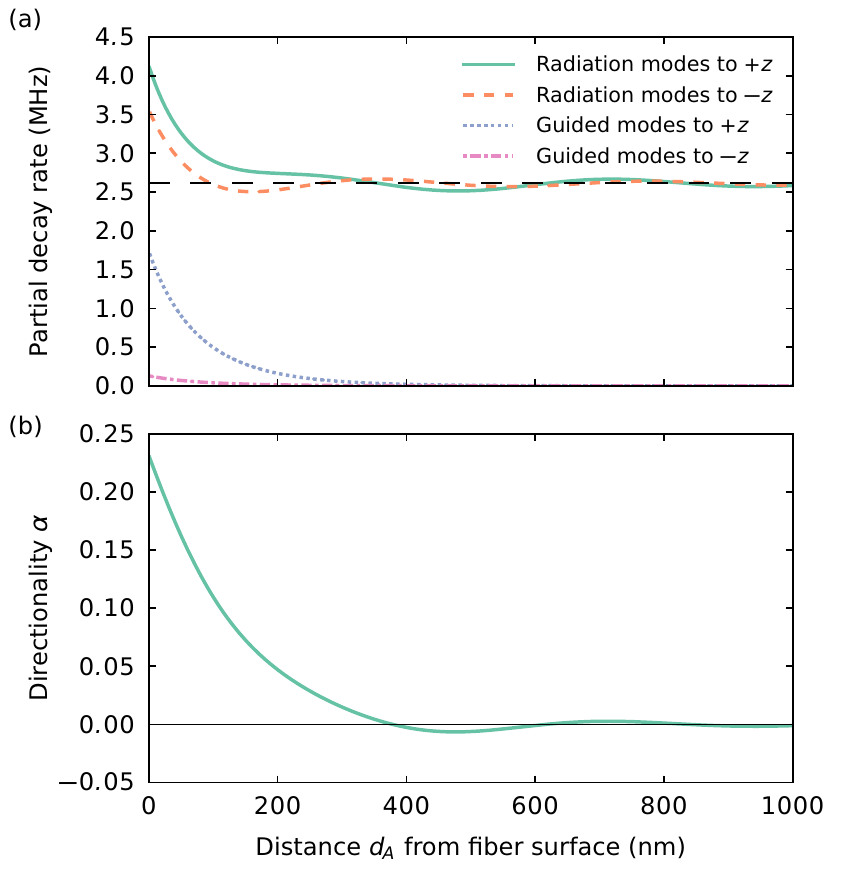}
\caption{(a) Partial decay rates $\gamma/2\pi$ into radiation and
  guided modes that propagate towards the $+z$ or $-z$ half space
  plotted as a function of the atom-surface distance $d_{\!A}$. The
  horizontal black dashed line indicates the free-space limit
  $\Gamma_\infty / 4\pi $.  (b)~Directionality $\alpha$ of the total
  spontaneous emission. For parameters see text and
  Fig.~\ref{fig:schematic}.}
\label{fig:DirEm}
\end{figure}

The dependence of the partial decay rates on the atom-fiber
distance $d_{\!A}$ is shown in Fig.~\ref{fig:DirEm}(a). The guided modes
are excited strongly asymmetrically, with emission to $+z$ being about
10 times more likely than to $-z$. This is the result of the inherent
link between local polarization and propagation direction in
strongly confined light fields~\cite{LeKien14a} and matches the observations of a
recent experiment~\cite{Mitsch14b}. As the intensity of the guided
nanofiber modes decreases with $d_{\!A}$, so does the fraction of
light that is coupled into them. The radiation modes show a complex
behavior: The excitation of radiation modes that propagate into the
$+z$ or $-z$ direction is, in general, also asymmetric.  Moreover, the
emission rates into these modes show characteristic Drexhage-type
oscillations~\cite{Drexhage70}. Remarkably, depending on the radial
distance of the emitter to the nanofiber surface, either the radiation
modes that propagate into the $+z$ or into the $-z$ direction are
excited more strongly. The amplitude of the oscillatory behavior
decreases with increasing emitter-surface distance. The radiative
emission into the $\pm z$ half-space is symmetric for emitters far away from
the nanofiber, i.e.~for ($d_{\!A}\gg \lambda_{10}$). In this limit,
the total emission rate of the atom approaches the free-space value of
$\Gamma_\infty/(2\pi)=5.234$~MHz~\cite{Steck10}.

For an atom in the proximity of the nanofiber, the combined emission
rates into the $\pm z$ half space are in general not equal. The
asymmetric emission should give rise to a force on the atom parallel
to the fiber axis. We quantify the asymmetry by introducing a
directionality parameter $\alpha$, which is a sum over all partial
decay rates weighted by the projection of the respective wave vector
onto the fiber axis. We denote the positive propagation constant of
the guided modes at frequency $\omega_{10}$ by $\beta_{10}$. The
weights are then $f \beta_{10}/k_{10}$ for guided
modes and $k_z/k_{10}$ for radiation modes. Finally, we make $\alpha$
dimensionless by normalizing to the total emission rate,
\begin{multline}
  \label{eq:directionality}
  \alpha(\vec{r}_{\!A}) = \frac{1}{\Gamma(\vec{r}_{\!A})}
  \bigg( \sum_{f,p} \frac{f \beta_{10}}{k_{10}}
  \gamma_{fp}^{\text{(G)}}(\vec{r}_{\!A}) \\
  + \sum_{m,p} \int_{-k_{10}}^{k_{10}} \!\! \dif k_z\,
  \frac{k_z}{k_{10}}\gamma_{k_z m p}^{\text{(R)}}(\vec{r}_{\!A})
  \bigg).
\end{multline}

Figure~\ref{fig:DirEm}(b) shows the directionality $\alpha$ as function of
$d_{\!A}$. An oscillatory, asymmetric emission is clearly visible. At
the surface, the directionality reaches more than 20~\%. The
directionality is independent of the magnitude of the dipole
moment. However, it strongly depends on the polarization of the
emitted light: For the emission of a $\sigma^-$ instead of a
$\sigma^+$-polarized photon, the directionality changes sign and,
thus, can be controlled via the internal state of the atom. The
emission is symmetric for the emission of $\pi$-polarized light.

The described directional emission in conjunction with the
conservation of total momentum in the system implies a lateral force
on the atom that points towards the
direction of stronger emission. This force can be viewed as a Casimir--Polder force on an atom at position $\vec{r}_{\!A}$. It is given by
the ensemble-averaged Lorentz force~\cite{Buhmann04}
$\vec{F}=\vec{\nabla}
\bigl\langle\hat{\vec{d}}\sprod\hat{\vec{E}}(\vec{r})\bigr\rangle
\big|_{\vec{r}=\vec{r}_{\!A}}$
of the quantized vacuum electric field $\hat{\vec{E}}$ acting on the atomic dipole moment $\hat{\vec{d}}$. After solving the coupled atom-field
dynamics and evaluating the averages over the atomic and field states,
one finds for an atom initially in an energy eigenstate $|n\rangle$ a
force
\mbox{$\vec{F}_n(\vec{r}_{\!A})
  =\vec{F}_n^\mathrm{nres}(\vec{r}_{\!A})+
  \vec{F}_n^\mathrm{res}(\vec{r}_{\!A})$}.
The Green's tensor can be decomposed into a vacuum part
$\tens{G}^{(0)}$ and a part $\tens{G}^{(1)}$ associated to the
scattering of the electric field from the nanofiber. Only the latter can give rise to a force. The non-resonant part
of the force, $\vec{F}_n^\text{nres}(\vec{r}_{\!A})$, is proportional
to the gradient of the \emph{symmetric} part of $\tens{G}^{(1)}$. The
resonant part, $\vec{F}_n^\text{res}(\vec{r}_{\!A})$, contains the
gradient of the \emph{Hermitian} part of $\tens{G}^{(1)}$~\cite{Supplement}.

When the atomic state $|n\rangle$ and the Hamiltonian are
time-reversal symmetric, the dipole-matrix elements can be chosen
real. In such cases, the gradient $\vec{\nabla}$ can be replaced
with a total derivative $\vec{\nabla}_{\vec{r}_{\!A}}$, showing that
no forces exist in a direction in which the system is translationally
invariant. We require an atom in an eigenstate that is \textit{not}
time-reversal symmetric, such that complex dipole-matrix elements can
give rise to a lateral force.

To see whether such forces can exist, let us consider an infinitely long
cylinder. Its Green's tensor is translationally invariant
along the cylinder axis $z$, so that
\mbox{$(\partial/\partial z)\tens{G}^{(1)}(\vec{r},\vec{r}',\omega)
=-(\partial/\partial z')\tens{G}^{(1)}(\vec{r},\vec{r}',\omega)$} holds.
Combining this with the Onsager reciprocity relation
\mbox{$\tens{G}^{(1)}(\vec{r},\vec{r}',\omega)
=\tens{G}^{(1)\trans}(\vec{r}',\vec{r},\omega)$}, we find that the
derivative with respect to $z$ of the scattering part of the Green's
tensor is anti-symmetric.
This immediately shows that for a ground-state atom with its purely
off-resonant CP force a lateral force cannot exist. Such a lateral force is also
forbidden by energy conservation: If it existed, one could use it to accelerate
a ground-state atom along the fiber and, thus, gain kinetic energy while
leaving the internal energy of the atom and that of the environment
unchanged.

A lateral force can thus only arise due to the resonant component
which is associated with the recoil of the atom when undergoing an
optical transition between two states. These forces are fueled by the
atom's internal energy. Using the anti-symmetry mentioned above, the
lateral force is given by
\begin{multline}
\label{SYB5}
F_{n,z}^\text{res}(\vec{r}_{\!A})
=2\mi\mu_0\sum_{k<n}\omega_{nk}^2\\
\times\vec{d}_{nk}\sprod\,\frac{\partial}{\partial z}
\operatorname{Im}\tens{G}^{(1)}(\vec{r},\vec{r}_{\!A},\omega_{nk})
\sprod\vec{d}_{kn}
\big|_{\vec{r}=\vec{r}_{\!A}}.
\end{multline}
Being proportional to $\operatorname{Im}\tens{G}$, it is associated
with the out-of-phase interaction of the electric-dipole oscillations with the reflected electric field. By contrast, the normal resonant CP force depends on the real part of the Green's tensor.

For the numerical evaluation of the force, we insert the scattering part of the
Green's tensor
\begin{multline}
\label{SYB6}
\tens{G}^{(1)}(\vec{r},\vec{r}',\omega)
=\frac{\mi}{8\pi}
\int_{-\infty}^\infty \dif k_z \sum_{m=0}^\infty
\sum_{p,p'=T\!E,T\!M}\\
\times\frac{2-\delta_{0m}}{k_\rho^2}
r_{pp'}
\vec{a}_{k_z mp}(\vec{r})
\otimes\vec{a}_{-k_z mp'}(\vec{r}')
\end{multline}
($\delta_{ij}$: Kronecker symbol) of a cylinder, which is given in
terms of cylindrical vector wave functions $\vec{a}_{k_z mp}$
and the respective reflection coefficients $r_{pp'}$~\cite{Li00} with
the dispersion relation $\omega^2/c^2=k^2=k_z^2+k_\rho^2$. Assuming a
cesium atom
in the excited state $|n\rangle=|1\rangle$ with a single downward
transition to
the ground state $|k\rangle=|0\rangle$, and using the complex refractive index
$1.45+\mi 2.05\times 10^{-7}$ of fused silica~\cite{Kitamura07} at the
transition frequency $\omega_{10}$, we can numerically evaluate the lateral
force (\ref{SYB5}). We compare this to a calculation based on the solution of
the Schrödinger equation where the expression for the lateral force
based on the guided and radiation modes of the nanofiber is~\cite{Supplement}
\begin{multline}
  \label{eq:lateral_force_SE}
  F_{1,z}(\vec{r}_{\!A}) = - \hbar
  \bigg( \sum_{f,p} f \beta_{10} \gamma_{fp}^{\text{(G)}}(\vec{r}_{\!A})\\
    + \sum_{m,p}\int_{-k_{10}}^{k_{10}} \!\dif k_z\, k_z
    \gamma_{k_z m p}^{\text{(R)}}(\vec{r}_{\!A})
  \bigg).
\end{multline}
Here, the recoil nature of the force becomes evident, as the
summation is over partial decay rates into guided and radiation modes multiplied
by the photon momentum of the respective mode.

\begin{figure}
\includegraphics[width=\columnwidth]{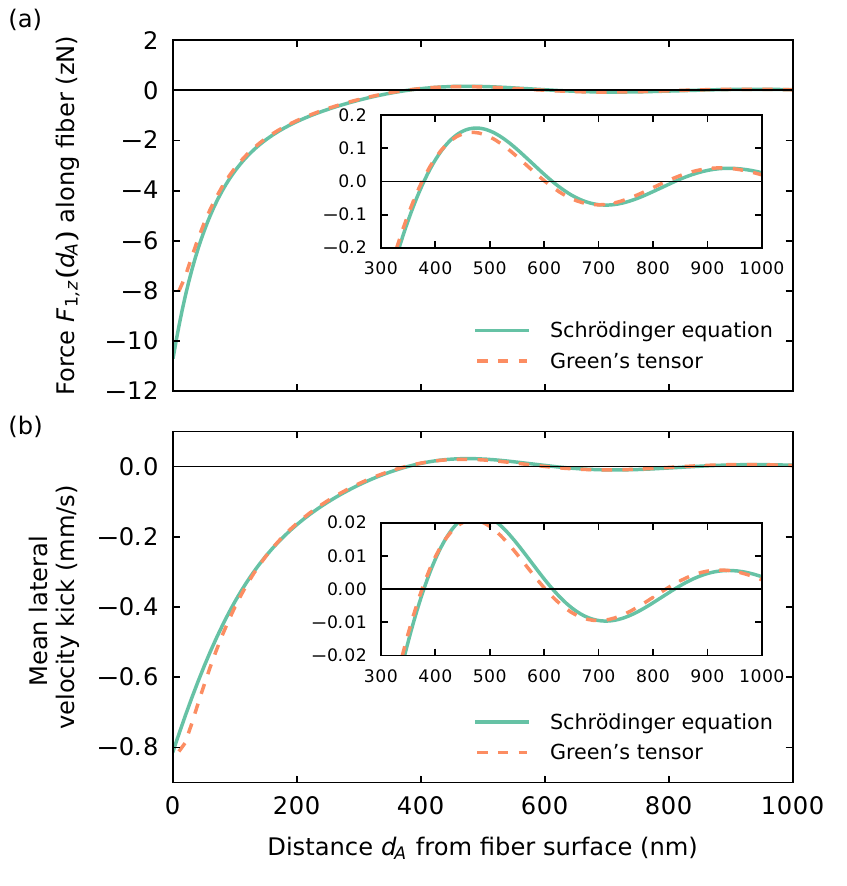}
\caption{(a) Lateral Casimir-Polder force $F_{1,z}(d_A,t=0)$ on an excited
cesium atom at distance as $d_{\!A}$ from the nanofiber
surface. (b)~Mean lateral velocity increase per emitted photon. Force and
velocity have been calculated using a method based on the solution of
the Schrödinger equation, and using the Green's tensor. Insets show
magnifications for distances above 300~nm.
We attribute the small deviation between the two methods to numerical
artifacts. For parameters see text.
}
\label{fig:force}
\end{figure}

The result of the numerical evaluation of Eqs.~\eqref{SYB5}
and~\eqref{eq:lateral_force_SE} is displayed in
Fig.~\ref{fig:force}(a). One observes an oscillating
force
with a distance dependence similar to the directionality parameter
shown in
Fig.~\ref{fig:DirEm}(b) but which, as expected from momentum conservation,
points into the direction opposite to the dominant emission. The lateral force
is an illustration of the subtle differences between Casimir--Polder force and
potential~\cite{Buhmann04}: It exists in spite of the $z$-independence
of the Casimir--Polder potential and hence cannot be derived from the
latter.

Recall that the lateral force is a result of the decay of the atom to its ground state and the associated emission of a photon. Thus, the time dependence of the ensemble-averaged force is given by
$F_{1,z}(\vec{r}_{\!A},t) = \me^{-\Gamma t} F_{1,z}(\vec{r}_{\!A})$.
We thus obtain an average momentum kick per emitted photon of
\begin{equation}
\Delta \langle p_z \rangle = F_{1,z}^\mathrm{res}(\vec{r}_{\!A}) /
\Gamma(\vec{r}_{\!A}) = -\alpha
\hbar k_{10},
\end{equation}
where $\alpha$ is the directionality defined in Eq.~(\ref{eq:directionality}).
Figure~\ref{fig:force}(b) shows the velocity
gained per photon, which can reach values of up to $0.8\mbox{ mm/s}$ in our situation.

The agreement between both methods provides some understanding of the
mechanism that underlies the lateral force. Note, however, that the
Green's tensor approach is more general and also allows for the investigation
of situations where the fiber shows substantial losses that prevent the
introduction of orthonormal modes. Indeed, an artificial increase of the
imaginary part of the refractive index by five orders of magnitude to $2\times
10^{-2}$ results in a 50~\% larger lateral force, which is likely to be driven by
the increased nonradiative decay that is automatically captured in the Green's
tensor formalism.

To gain some further intuition into the lateral force, we consider the
simpler
case of an atom in front of a semi-infinite half space. In this case, the
scattering Green tensor takes the form
\begin{multline}
\label{B97}
\tens{G}^{(1)}(\vec{r},\vec{r}',\omega)
=\frac{\mi}{8\pi^2}\,\int\frac{\dif^2k^\parallel}
 {k_x}\,
 \me^{\mi\vec{k}^\parallel\cdot(\vec{r}-\vec{r}')
  +\mi k_x(z+z')}\\
\times\sum_{p=T\!E,T\!M}r_p
 \vec{e}_{\sigma+}\tprod
 \vec{e}_{\sigma-}
\end{multline}
with polarization unit vectors $\vec{e}_{\sigma\pm}$, Fresnel
reflection coefficients $r_p$ and the dispersion relation
$\omega^2/c^2=k^2=k_x^2+{(k^{\parallel})}^2$. Substitution into
Eq.~(\ref{SYB5}) then leads to a simple expression for the lateral
Casimir--Polder force in the retarded limit $d_A\gg\lambda_{10}$: in
this case, the wave-vector integral is dominated by the
stationary-phase point at $k^\parallel=0$ and we find for weak
absorption:
\begin{equation}
F_{1,z}^\mathrm{res}(d_{\!A})
=\frac{\operatorname{Im}(d_{01,z}d_{10,x})\omega_{10}^2}
{8\varepsilon_0c^2 d_A^2}\sin(2d_{\!A} k_{10})\,r_p
\end{equation}
with $r_p=[n(\omega_{10})-1]/[n(\omega_{10})+1]$. One sees that the
lateral force is due to the interaction of the atom with its image dipole behind the surface of the half space. Also for this model case, we observe Drexhage-type oscillations.

In summary, we have theoretically described a translationally invariant
lateral scattering force which arises from asymmetric spontaneous emission of a
circular dipole emitter that is in close proximity to an optical nanofiber. In
contrast to all lateral forces previously studied in Casimir and
Casimir--Polder physics, the force does not rely on a corrugation of the
material surface. Moreover, the magnitude and sign of the force depend on the
polarization of the emitted light and thus can be controlled, e.g., by the
quantum state of an atomic emitter. The described lateral force is generic
in the sense that it prevails also in other geometries, e.g. for circular
emitters above a plane surface.
In contrast to the forces on Mie particles proposed in \cite{Bliokh14}, our
lateral force does not rely on a
higher-order interaction between electric- and magnetic-induced dipoles.

In optical force measurements on particles on a substrate~\cite{Neuman04}, this effect
will influence measurement outcomes as soon as scattering becomes relevant.
Moreover, we expect the lateral force to enrich the dynamics of
optically driven self-organization of atomic ensembles close to
waveguides~\cite{Chang13,Holzmann14}. In order to tune the force, the
atom can be coupled to, e.g., a whispering-gallery mode optical
resonator~\cite{Junge13,Shomroni14}, thereby changing the relative
share of emission into guided and radiative modes. Lateral
forces as described here will also influence laser cooling of atoms close
to surfaces~\cite{Hammes03,Stehle14} and nanophotonic
structures~\cite{Vetsch12,Thompson13b,Goban14}. The study of lateral
scattering forces might be extended to thermal Casimir--Polder forces
or to other geometries, e.g., an atom above a sphere where the lateral
position is coupled to the atom-surface distance.

We thank F.~Ciccarello, F.~Intravaia, Fam Le Kien, A.~Rauschenbeutel, C.~Sayrin, and J.~Volz for helpful comments and discussions. Financial support by the DFG (grant no. SCHE 612/2-1) is gratefully
acknowledged. S.~Y.~B.~gratefully acknowledges support by the DFG (grant BU
1803/3-1) and the Freiburg Institute for
Advanced Studies.

\appendix

\section{Partial emission rates}

In this Supplement we present the calculation based on the solution of
the Schrödinger equation for a cesium atom close to the nanofiber. The
atom is initially in the hyperfine state
$\ket{1} = \ket{F'=5, M_F'=5}$ of the $6P_{3/2}$ manifold. The only
available decay channel is to the hyperfine ground state
$\ket{0} = \ket{F=4, M_F=4}$ through the emission of a
$\sigma^+$-polarized photon at frequency $\omega_{10}$, and we can
treat the atom as an effective two-level system. In the interaction
picture, the atomic dipole operator is given by
\begin{equation}
  \label{eq:dipole_operator}
  \hat{\vec{d}} = \vec{d}_{01} \ket{0}\!\!\bra{1}e^{-\mi
    \omega_{10} t} + \vec{d}_{10} \ket{1}\!\!\bra{0}e^{\mi
    \omega_{10} t},
\end{equation}
where $\vec{d}_{01} = \vec{d}_{10}^*$ is the corresponding dipole matrix
element. We follow~\cite{LeKien05c} and decompose the electric field into
contributions from guided and radiation modes, neglecting material
absorption. For the guided modes we assume that the single-mode
condition is satisfied for a finite bandwidth around $\omega_{10}$. In
cylindrical coordinates $\vec{r} = (r, \phi, z)$, the positive-frequency part of
the guided-mode field operator in the interaction picture can be written as
\begin{multline}
  \label{eq:field_guided}
  \hat{\vec{E}}_G^{(+)}(\vec{r}) = \mi \sum_{f, p}\int_0^\infty \!\!\dif
  \omega \sqrt{\frac{\hbar \omega \beta'}{4 \pi \epsilon_0}}
    \\  \times \hat{a}_{\omega f p} \vec{e}_{\omega f p}(r)
     e^{-\mi(\omega t - f \beta z - p \phi)}.
\end{multline}
Here, $\beta$ is the propagation constant of the guided mode, $\beta'
= \dif \beta/\dif \omega$, $\hat{a}_{\omega f p}$ is the annihilation
operator and $\vec{e}_{\omega f p}$ is the profile function of the
guided mode. The indices $f=\pm 1$ and $p=\pm 1$ indicate the
propagation direction and the handedness of the quasi-circular
polarization, respectively~\cite{LeKien05c}. Similarly, for the
radiation modes
\begin{multline}
  \label{eq:field_radiation}
  \hat{\vec{E}}_R^{(+)}(\vec{r}) = \mi \sum_{m, p}\int_0^\infty
  \!\!\dif \omega
  \int_{-\omega/c}^{\omega/c} \!\! \dif k_z \sqrt{\frac{\hbar \omega}{4
      \pi \epsilon_0}}
  \\ \times \hat{a}_{\omega k_z m p} \vec{e}_{\omega k_z m p}(r)
    e^{-\mi(\omega t - k_z z - m \phi)},
\end{multline}
where $k_z$ is the projection of the wave vector onto the fiber axis,
$m \in \mathbb{Z}$ is the mode order, and $p$ is the mode polarization.

Let $\vec{r}_A$ denote the position of the atom. The Hamiltonian for
the atom-field interaction in the dipole and rotating-wave
approximations is given by
\begin{equation}
\label{eq:hamiltonian}
\begin{split}
  \hat{H}_\text{int} &= -\hat{\vec{d}} \cdot \hat{\vec{E}}(\vec{r}_A) \\
  &=-\mi \hbar \sum_{f,p} \int_0^\infty \!\! \dif \omega\, G_{\omega f
    p}(\vec{r}_A) \ket{1}\!\!\bra{0} \hat{a}_{\omega f p}
  e^{-\mi(\omega-\omega_{10})t} \\
  & \quad -\mi \hbar \sum_{m,p} \int_0^\infty \!\! \dif \omega
  \int_{-\omega/c}^{\omega/c} \!\! \dif k_z \, G_{\omega k_z m
    p}(\vec{r}_A) \\
  & \qquad  \times \ket{1}\!\!\bra{0} \hat{a}_{\omega k_z m p}
  e^{-\mi(\omega-\omega_{10})t} + \text{H. c.}
\end{split}
\end{equation}
We have introduced the coefficients $G_{\omega f p}(\vec{r})$ and
$G_{\omega k_z m p}(\vec{r})$ which characterize the coupling of
the atomic transition to a specific guided or radiation mode, respectively.
These coefficients are given by
\begin{equation}
\label{eq:couplings}
\begin{split}
G_{\omega f p}(\vec{r}) &= \sqrt{\frac{\omega \beta'}{4\pi\epsilon_0 \hbar}}
\vec{d}_{10} \cdot \vec{e}_{\omega f p}(r)\, e^{\mi(f \beta z + p \phi)},
\\
G_{\omega k_z m p}(\vec{r}) &= \sqrt{\frac{\omega}{4 \pi \epsilon_0
    \hbar}}
\vec{d}_{10} \cdot \vec{e}_{\omega k_z f p}(r)\, e^{\mi(k_z z + m
  \phi)}.
\end{split}
\end{equation}

For the combined atom-field state, we restrict the Hilbert space to a single excitation. In the following, the first element of the state vector indicates the state of the atom and the second element that of the field. Then, the atom-field state at any time can be written as
\begin{gather}
\ket{\psi(t)} = c_1(t)\ket{1, 0} + \sum_{f, p}\int_0^\infty \!\! \dif
\omega\, c_{\omega f p}(t) \ket{0, 1_{\omega f p}} \nonumber \\
+ \sum_{m,p} \int_0^\infty \!\! \dif \omega
\int_{-\omega/c}^{\omega/c} \!\! \dif k_z \, c_{\omega k_z m
  p}\ket{0, 1_{\omega k_z m p}}.
\label{eq:state}
\end{gather}
We obtain expressions for the time-dependent coefficients $c(t)$ by
inserting into the Schrödinger equation and following a standard
Wigner-Weisskopf treatment. First, we formally integrate the
differential equations for the guided and radiation mode
coefficients. Applying a Markov approximation, we replace $c_1(t')$ by
$c_1(t)$ in the resulting expressions and obtain
\begin{equation}
  \label{eq:field_coefficients}
  \begin{split}
  c_{\omega f p}(t) &= c_1(t)\, G_{\omega f p}^*(\vec{r}_A) \int_0^t
  \!\! \dif t' e^{\mi(\omega - \omega_{10})t'}, \\
  c_{\omega k_z m p}(t) &= c_1(t)\, G_{\omega k_z m p}^*(\vec{r}_A) \int_0^t
  \!\! \dif t' e^{\mi(\omega - \omega_{10})t'}. \\
\end{split}
\end{equation}
These are inserted into the differential equation for $c_1(t)$, where
we let the upper limits of the time integrals tend to
infinity. Solving the integrals, we further neglect the term
corresponding to a vacuum frequency shift and obtain a simple
exponential decay, $c_1(t) = e^{-\Gamma t/2}$ with a total,
position-dependent spontaneous-emission rate $\Gamma(\vec{r}_{\!A})$ as
\begin{multline}
\label{eq:gamma}
\frac{\Gamma(\vec{r}_{\!A})}{2\pi} =
\sum_{f,p}\left|G_{\omega_{10}fp}(\vec{r}_A)\right|^2  \\
  + \sum_{m,p}\int_{-k_{10}}^{k_{10}} \!\! \dif k_z\, \left|G_{\omega_{10} k_z
    m p}(\vec{r}_A)\right|^2.
\end{multline}
We see that each guided mode has an associated partial
emission rate given by
\begin{equation}
  \label{eq:guided_rates}
  \begin{split}
    \gamma_{fp}^{\text{(G)}}(\vec{r}_{\!A}) &= 2 \pi \left| G_{\omega_{10} f
        p}(\vec{r_{\!A}}) \right|^2 \\
    &= \frac{\omega_{10} \beta_{10}'}{2 \epsilon_0 \hbar} \left |
      \vec{d}_{10} \cdot \vec{e}_{\omega_{10}
        fp}(\vec{r}_{\!A})\right|^2,
  \end{split}
\end{equation}
and similarly for the radiation modes,
\begin{equation}
  \label{eq:radiation_rates}
  \begin{split}
    \gamma_{k_z mp}^{\text{(R)}}(\vec{r}_{\!A}) &= 2 \pi \left|
      G_{\omega_{10} k_z mp}(\vec{r_{\!A}}) \right|^2 \\
    &= \frac{\omega_{10}}{2 \epsilon_0 \hbar} \left |
      \vec{d}_{10} \cdot \vec{e}_{\omega_{10} k_z
        mp}(\vec{r}_{\!A})\right|^2.
  \end{split}
\end{equation}

To study the directional dependence of the spontaneous emission, we
note that the partial decay rates into each guided or radiation mode
are proportional to the absolute square of the coupling coefficients
\eqref{eq:couplings}. The propagation direction along the fiber axis
for the guided modes is encoded in the parameter $f$, such that $f=+1$
corresponds to emission into the positive half space. Similarly, for
the radiation modes it is encoded into the projection $k_z$ of the
wave vector, and emission into the positive half space is represented
by modes with $k_z > 0$. Hence, the partial decay rates into the half
spaces $\pm z$ for guided ($\gamma_{\pm}^\text{(G)}$) and
radiation modes ($\gamma_{\pm}^\text{(G)}$) are given by
\begin{equation}
  \label{eq:partial_rates}
  \begin{split}
    \gamma_{\pm}^\text{(G)}(\vec{r}_{\!A}) &=  \sum_p \gamma_{f=\pm 1,p}^{\text{(G)}}(\vec{r}_{\!A}),\\
    \gamma_{\pm}^\text{(R)}(\vec{r}_{\!A}) &= \pm \sum_{m,p}
    \int_0^{\pm k_{10}} \!\! \dif k_z\,
    \gamma_{k_z mp}^{\text{(R)}}(\vec{r}_{\!A}),\\
  \end{split}
\end{equation}
which is Eq.~(2) in the main text.

\section{Lorentz Force}
The calculation of the Lorentz force amounts to computing the expectation
value $\langle \vec{F} \rangle = \vec{\nabla} \langle \hat{\vec{d}} \cdot
\hat{\vec{E}}(\vec{r}) \rangle |_{\vec{r} = \vec{r}_A}$. We note that, in
contrast to the interaction Hamiltonian~\eqref{eq:hamiltonian}, here the field
operator is evaluated at a general position $\vec{r}$. The position of
the atom is only inserted after taking the gradient.

We are only interested in the lateral part of the force, i.e., $\langle
F_z \rangle = \frac{\partial}{\partial z}\langle \hat{\vec{d}} \cdot
\hat{\vec{E}}(\vec{r}) \rangle|_{\vec{r}=\vec{r}_A}$. To keep track of
the individual contributions to the force, we label the part of the
state~\eqref{eq:state} that describes excitations of the guided modes
$\ket{\psi_G}$, and similarly for the radiation modes
$\ket{\psi_R}$. The Lorentz force is then given by the sum of
two terms, $\langle F_z \rangle = \langle F_z^\text{(G)} \rangle +
\langle F_z^\text{(R)} \rangle $, where
\begin{equation}
\label{eq:force_reduced}
  \langle F_z^\text{(M)} \rangle = 2 \operatorname{Re} \left(
      c_1^*(t) \frac{\partial}{\partial z} V_{1,M}(\vec{r})\right)
    \bigg|_{\vec{r}=\vec{r}_A},
\end{equation}
and $M \in \{G, R\}$. We have introduced the matrix element $V_{1,M} =
\bra{1,0} \hat{\vec{d}}^{(-)} \cdot
\hat{\vec{E}}_M^{(+)}(\vec{r})\ket{\psi_M}$, where
$\hat{\vec{d}}^{(-)}$ is the negative-frequency part of the dipole
operator, i.e., the second term in Eq.~\eqref{eq:dipole_operator}.

For the guided modes, we find
\begin{equation}
  \label{eq:Veg}
  \begin{split}
    V_{1,G}(\vec{r}) &= - \hbar e^{-\Gamma t/2} \sum_{f,p} \int_0^\infty \dif
    \omega\, G_{\omega f p}(\vec{r}) G_{\omega f p}^*(\vec{r}_A) \\
    &\qquad \times \frac{e^{-\mi(\omega - \omega_{10})t}-1}{\omega
      - \omega_{10}} \\
    & \approx - A(t) e^{-\Gamma t/2}
    \frac{\beta_{10}'}{4\pi\epsilon_0} \\
    & \qquad \times \sum_{f,p}
    [\vec{d}_{10}\cdot \vec{e}_{\omega_{10}fp}(\vec{r})]
    {[\vec{d}_{10}\cdot \vec{e}_{\omega_{10}fp}(\vec{r}_{\!A})]}^* \\
    & \qquad \qquad \times
    e^{\mi[f\beta_{10}(z-z_A) + p(\phi-\phi_A)]},
  \end{split}
\end{equation}
where for the approximation we have made use of the fact that the mode
functions, propagation constant, and their derivatives vary slowly in a
frequency interval around $\omega_{10}$, for which we expect the main
contribution to the integral in~\eqref{eq:Veg}. We hence replace them
by their resonant values, such that the remaining integral is
\begin{equation}
\label{eq:Aoft}
\begin{split}
  A(t) &= \int_0^\infty \!\! \dif \omega \, \omega \frac{e^{-\mi(\omega
      - \omega_{10})t}-1}{\omega - \omega_{10}} \\
  &\approx \int_{-\infty}^\infty \!\! \dif \Delta\, (\omega_{10}
  + \Delta)\frac{e^{-\mi \Delta t} - 1}{\Delta}
  = -\mi \pi \omega_{10}.
\end{split}
\end{equation}
In the second line, we substituted $\Delta = \omega - \omega_{10}$ and
extended the lower limit of the integral from $-\omega_{10}$ to
$-\infty$.
Inserting~\eqref{eq:Aoft} and~\eqref{eq:Veg}
into~\eqref{eq:force_reduced}, we find that the lateral force due to
spontaneous emission into guided modes is
\begin{equation}
  \label{eq:Fz,G}
  \langle F_z^\text{(G)} \rangle = -e^{-\Gamma t} \frac{\omega_{10}
    \beta_{10} \beta_{10}'}{2 \epsilon_0} \sum_{f,p} f \left|
      \vec{d}_{10} \cdot \vec{e}_{\omega_{10}fp}(\vec{r}_{\!A})\right|^2.
\end{equation}

The calculation for the contribution from the radiation modes is
similar. The approximation that the main contribution stems from a
narrow frequency interval around the resonance allows us to set
the limits of the integral over $k_z$ to $\pm k_{10}$ and take it out
of the frequency integral. Hence,
\begin{gather}
  V_{1,R}(\vec{r}) \approx - \frac{A(t) e^{-\Gamma t/2}}{4 \pi
    \epsilon_0} \sum_{m,p} \int_{-k_{10}}^{k_{10}} \!\! \dif k_z
e^{\mi[k_z(z-z_A) + m(\phi-\phi_A)]} \nonumber \\
  \times
    [\vec{d}_{10}\cdot \vec{e}_{\omega_{10}k_z mp}(r)]
    {[\vec{d}_{10}\cdot \vec{e}_{\omega_{10}k_z mp}(\vec{r}_{\!A})]}^*
\label{eq:Ver}
\end{gather}
such that
\begin{equation}
  \label{eq:Fz,R}
  \langle F_z^\text{(R)} \rangle = -e^{-\Gamma t} \frac{\omega_{10}}{2
    \epsilon_0} \sum_{m,p} \int_{-k_{10}}^{k_{10}} \!\! \dif k_z \, k_z \left|
      \vec{d}_{10} \cdot \vec{e}_{\omega_{10}k_z mp}(\vec{r}_{\!A})\right|^2.
\end{equation}

Finally, we note that Eqs.~\eqref{eq:Fz,G} and~\eqref{eq:Fz,R} can be
rewritten in terms of the partial decay rates of
Eqs.~\eqref{eq:guided_rates} and~\eqref{eq:radiation_rates},
\begin{equation}
  \label{eq:force_with_rates}
  \begin{split}
    \langle F_z^{\text{(G)}} \rangle &= - \hbar e^{-\Gamma t}
    \sum_{f,p} f \beta_{10} \gamma_{fp}^{\text{(G)}}(\vec{r}_{\!A}), \\
    \langle F_z^{\text{(R)}} \rangle &= - \hbar e^{-\Gamma t}
    \sum_{m,p} \int_{-k_{10}}^{k_{10}} \! \dif k_z \, k_z \gamma_{k_z
      mp}^{\text{(R)}}(\vec{r}_{\!A}).
  \end{split}
\end{equation}

\section{Resonant and non-resonant parts of the Lorentz force}
In general, the Lorentz force contains contributions from resonant and
non-resonant components, \mbox{$\vec{F}_n(\vec{r}_{\!A})
  =\vec{F}_n^\mathrm{nres}(\vec{r}_{\!A}) \rangle +
  \vec{F}_n^\mathrm{res}(\vec{r}_{\!A})$}. We give here their full
analytical expressions based on the Green's tensor $\tens{G}$. In particular, the non-resonant component is~\cite{Buhmann04}
\begin{multline}
\label{SYB2}
\vec{F}_n^\mathrm{nres}(\vec{r}_{\!A})
=-\frac{2\mu_0}{\pi}\sum_k
\int_0^\infty \dif\xi\,\frac{\xi^2\omega_{kn}}{\omega_{kn}^2+\xi^2}\\
\times\vec{\nabla}
\vec{d}_{nk}\sprod\mathcal{S}
\tens{G}^{(1)}(\vec{r},\vec{r}_{\!A},\mi\xi)\sprod
\mathbf{d}_{kn}\big|_{\vec{r}=\vec{r}_A}~,
\end{multline}
where \mbox{$\mathcal{S}\tens{G}=(\tens{G}+\tens{G}^\trans)/2$} is the symmetric part of a tensor. The resonant component is given by
\begin{multline}
\label{SYB3}
\vec{F}_n^\mathrm{res}(\vec{r}_{\!A})
= 2\mu_0\sum_{k<n}\omega_{nk}^2\\
\times\vec{\nabla}\vec{d}_{nk}\sprod
\mathcal{H}\tens{G}^{(1)}(\vec{r},\vec{r}_{\!A},\omega_{nk})
\sprod\vec{d}_{kn}
\big|_{\vec{r}=\vec{r}_{\!A}}~,
\end{multline}
where \mbox{$\mathcal{H}\tens{G}=(\tens{G}+\tens{G}^{\ast\trans})/2$} is the
Hermitian part. Here, $\omega_{kn}$ and $\vec{d}_{nk}$ are the
frequencies and matrix elements for electric-dipole transitions of
the atom and $\tens{G}^{(1)}$ is the scattering part of the classical Green's
tensor for the electric field.

The directionality parameter $\alpha$ given in the main manuscript can
also be expressed in terms of the Green's tensor as
\begin{equation}
\alpha =
\frac{-2}{k_{10}}
\left[ \frac{\partial}{\partial z}
\operatorname{Im}G_{xz}^{(1)}\right] \bigg[ \operatorname{Im}G_{xx}
+\operatorname{Im}G_{zz} \bigg]^{-1}.
\end{equation}

\bibliography{Lateral}

\begin{thebibliography}{41}%
\makeatletter
\providecommand \@ifxundefined [1]{%
 \@ifx{#1\undefined}
}%
\providecommand \@ifnum [1]{%
 \ifnum #1\expandafter \@firstoftwo
 \else \expandafter \@secondoftwo
 \fi
}%
\providecommand \@ifx [1]{%
 \ifx #1\expandafter \@firstoftwo
 \else \expandafter \@secondoftwo
 \fi
}%
\providecommand \natexlab [1]{#1}%
\providecommand \enquote  [1]{``#1''}%
\providecommand \bibnamefont  [1]{#1}%
\providecommand \bibfnamefont [1]{#1}%
\providecommand \citenamefont [1]{#1}%
\providecommand \href@noop [0]{\@secondoftwo}%
\providecommand \href [0]{\begingroup \@sanitize@url \@href}%
\providecommand \@href[1]{\@@startlink{#1}\@@href}%
\providecommand \@@href[1]{\endgroup#1\@@endlink}%
\providecommand \@sanitize@url [0]{\catcode `\\12\catcode `\$12\catcode
  `\&12\catcode `\#12\catcode `\^12\catcode `\_12\catcode `\%12\relax}%
\providecommand \@@startlink[1]{}%
\providecommand \@@endlink[0]{}%
\providecommand \url  [0]{\begingroup\@sanitize@url \@url }%
\providecommand \@url [1]{\endgroup\@href {#1}{\urlprefix }}%
\providecommand \urlprefix  [0]{URL }%
\providecommand \Eprint [0]{\href }%
\providecommand \doibase [0]{http://dx.doi.org/}%
\providecommand \selectlanguage [0]{\@gobble}%
\providecommand \bibinfo  [0]{\@secondoftwo}%
\providecommand \bibfield  [0]{\@secondoftwo}%
\providecommand \translation [1]{[#1]}%
\providecommand \BibitemOpen [0]{}%
\providecommand \bibitemStop [0]{}%
\providecommand \bibitemNoStop [0]{.\EOS\space}%
\providecommand \EOS [0]{\spacefactor3000\relax}%
\providecommand \BibitemShut  [1]{\csname bibitem#1\endcsname}%
\let\auto@bib@innerbib\@empty
\bibitem [{\citenamefont {Buhmann}(2012)}]{Buhmann12}%
  \BibitemOpen
  \bibfield  {author} {\bibinfo {author} {\bibfnamefont {S.~Y.}\ \bibnamefont
  {Buhmann}},\ }\href@noop {} {\emph {\bibinfo {title} {Dispersion Forces I ---
  Macroscopic Quantum Electrodynamics and Ground-State Casimir, Casimir--Polder
  and van der Waals Forces}}}\ (\bibinfo  {publisher} {Springer},\ \bibinfo
  {year} {2012})\BibitemShut {NoStop}%
\bibitem [{\citenamefont {{Wang}}\ and\ \citenamefont {{Chan}}(2014)}]{Wang14}%
  \BibitemOpen
  \bibfield  {author} {\bibinfo {author} {\bibfnamefont {S.~B.}\ \bibnamefont
  {{Wang}}}\ and\ \bibinfo {author} {\bibfnamefont {C.~T.}\ \bibnamefont
  {{Chan}}},\ }\href@noop {} {\bibfield  {journal} {\bibinfo  {journal} {Nat.
  Commun.}\ }\textbf {\bibinfo {volume} {5}},\ \bibinfo {eid} {3307} (\bibinfo
  {year} {2014})}\BibitemShut {NoStop}%
\bibitem [{\citenamefont {{Bliokh}}\ \emph {et~al.}(2014)\citenamefont
  {{Bliokh}}, \citenamefont {{Bekshaev}},\ and\ \citenamefont
  {{Nori}}}]{Bliokh14}%
  \BibitemOpen
  \bibfield  {author} {\bibinfo {author} {\bibfnamefont {K.~Y.}\ \bibnamefont
  {{Bliokh}}}, \bibinfo {author} {\bibfnamefont {A.~Y.}\ \bibnamefont
  {{Bekshaev}}}, \ and\ \bibinfo {author} {\bibfnamefont {F.}~\bibnamefont
  {{Nori}}},\ }\href@noop {} {\bibfield  {journal} {\bibinfo  {journal} {Nat.
  Commun.}\ }\textbf {\bibinfo {volume} {5}},\ \bibinfo {eid} {3300} (\bibinfo
  {year} {2014})}\BibitemShut {NoStop}%
\bibitem [{\citenamefont {{Hayat}}\ \emph {et~al.}(2014)\citenamefont
  {{Hayat}}, \citenamefont {{Balthasar M{\"u}ller}},\ and\ \citenamefont
  {{Capasso}}}]{ArXiv_Hayat14}%
  \BibitemOpen
  \bibfield  {author} {\bibinfo {author} {\bibfnamefont {A.}~\bibnamefont
  {{Hayat}}}, \bibinfo {author} {\bibfnamefont {J.~P.}\ \bibnamefont
  {{Balthasar M{\"u}ller}}}, \ and\ \bibinfo {author} {\bibfnamefont
  {F.}~\bibnamefont {{Capasso}}},\ }\href@noop {} {\  (\bibinfo {year}
  {2014})},\ \Eprint {http://arxiv.org/abs/1408.2268} {arXiv:1408.2268}
  \BibitemShut {NoStop}%
\bibitem [{\citenamefont {{Lin}}\ \emph {et~al.}(2013)\citenamefont {{Lin}},
  \citenamefont {{Mueller}}, \citenamefont {{Wang}}, \citenamefont {{Yuan}},
  \citenamefont {{Antoniou}}, \citenamefont {{Yuan}},\ and\ \citenamefont
  {{Capasso}}}]{Lin13}%
  \BibitemOpen
  \bibfield  {author} {\bibinfo {author} {\bibfnamefont {J.}~\bibnamefont
  {{Lin}}}, \bibinfo {author} {\bibfnamefont {J.~P.~B.}\ \bibnamefont
  {{Mueller}}}, \bibinfo {author} {\bibfnamefont {Q.}~\bibnamefont {{Wang}}},
  \bibinfo {author} {\bibfnamefont {G.}~\bibnamefont {{Yuan}}}, \bibinfo
  {author} {\bibfnamefont {N.}~\bibnamefont {{Antoniou}}}, \bibinfo {author}
  {\bibfnamefont {X.-C.}\ \bibnamefont {{Yuan}}}, \ and\ \bibinfo {author}
  {\bibfnamefont {F.}~\bibnamefont {{Capasso}}},\ }\href {\doibase
  10.1126/science.1233746} {\bibfield  {journal} {\bibinfo  {journal}
  {Science}\ }\textbf {\bibinfo {volume} {340}},\ \bibinfo {pages} {331}
  (\bibinfo {year} {2013})}\BibitemShut {NoStop}%
\bibitem [{\citenamefont {{Rodriguez-Fortuno}}\ \emph
  {et~al.}(2013)\citenamefont {{Rodriguez-Fortuno}}, \citenamefont {{Marino}},
  \citenamefont {{Ginzburg}}, \citenamefont {{O'Connor}}, \citenamefont
  {{Mart{\'{\i}}nez}}, \citenamefont {{Wurtz}},\ and\ \citenamefont
  {{Zayats}}}]{Rodriguez-Fortuno13}%
  \BibitemOpen
  \bibfield  {author} {\bibinfo {author} {\bibfnamefont {F.~J.}\ \bibnamefont
  {{Rodriguez-Fortuno}}}, \bibinfo {author} {\bibfnamefont {G.}~\bibnamefont
  {{Marino}}}, \bibinfo {author} {\bibfnamefont {P.}~\bibnamefont
  {{Ginzburg}}}, \bibinfo {author} {\bibfnamefont {D.}~\bibnamefont
  {{O'Connor}}}, \bibinfo {author} {\bibfnamefont {A.}~\bibnamefont
  {{Mart{\'{\i}}nez}}}, \bibinfo {author} {\bibfnamefont {G.~A.}\ \bibnamefont
  {{Wurtz}}}, \ and\ \bibinfo {author} {\bibfnamefont {A.~V.}\ \bibnamefont
  {{Zayats}}},\ }\href {\doibase 10.1126/science.1233739} {\bibfield  {journal}
  {\bibinfo  {journal} {Science}\ }\textbf {\bibinfo {volume} {340}},\ \bibinfo
  {pages} {328} (\bibinfo {year} {2013})}\BibitemShut {NoStop}%
\bibitem [{\citenamefont {Luxmoore}\ \emph {et~al.}(2013)\citenamefont
  {Luxmoore}, \citenamefont {Wasley}, \citenamefont {Ramsay}, \citenamefont
  {Thijssen}, \citenamefont {Oulton}, \citenamefont {Hugues}, \citenamefont
  {Fox},\ and\ \citenamefont {Skolnick}}]{Luxmoore13a}%
  \BibitemOpen
  \bibfield  {author} {\bibinfo {author} {\bibfnamefont {I.~J.}\ \bibnamefont
  {Luxmoore}}, \bibinfo {author} {\bibfnamefont {N.~A.}\ \bibnamefont
  {Wasley}}, \bibinfo {author} {\bibfnamefont {A.~J.}\ \bibnamefont {Ramsay}},
  \bibinfo {author} {\bibfnamefont {A.~C.~T.}\ \bibnamefont {Thijssen}},
  \bibinfo {author} {\bibfnamefont {R.}~\bibnamefont {Oulton}}, \bibinfo
  {author} {\bibfnamefont {M.}~\bibnamefont {Hugues}}, \bibinfo {author}
  {\bibfnamefont {A.~M.}\ \bibnamefont {Fox}}, \ and\ \bibinfo {author}
  {\bibfnamefont {M.~S.}\ \bibnamefont {Skolnick}},\ }\href {\doibase
  http://dx.doi.org/10.1063/1.4845975} {\bibfield  {journal} {\bibinfo
  {journal} {Appl. Phys. Lett.}\ }\textbf {\bibinfo {volume} {103}},\ \bibinfo
  {eid} {241102} (\bibinfo {year} {2013})}\BibitemShut {NoStop}%
\bibitem [{\citenamefont {Neugebauer}\ \emph {et~al.}(2014)\citenamefont
  {Neugebauer}, \citenamefont {Bauer}, \citenamefont {Banzer},\ and\
  \citenamefont {Leuchs}}]{Neugebauer14}%
  \BibitemOpen
  \bibfield  {author} {\bibinfo {author} {\bibfnamefont {M.}~\bibnamefont
  {Neugebauer}}, \bibinfo {author} {\bibfnamefont {T.}~\bibnamefont {Bauer}},
  \bibinfo {author} {\bibfnamefont {P.}~\bibnamefont {Banzer}}, \ and\ \bibinfo
  {author} {\bibfnamefont {G.}~\bibnamefont {Leuchs}},\ }\href@noop {}
  {\bibfield  {journal} {\bibinfo  {journal} {Nano Lett.}\ }\textbf {\bibinfo
  {volume} {14}},\ \bibinfo {pages} {2546} (\bibinfo {year}
  {2014})}\BibitemShut {NoStop}%
\bibitem [{\citenamefont {Mitsch}\ \emph {et~al.}(2014)\citenamefont {Mitsch},
  \citenamefont {Sayrin}, \citenamefont {Albrecht}, \citenamefont
  {Schneeweiss},\ and\ \citenamefont {Rauschenbeutel}}]{Mitsch14b}%
  \BibitemOpen
  \bibfield  {author} {\bibinfo {author} {\bibfnamefont {R.}~\bibnamefont
  {Mitsch}}, \bibinfo {author} {\bibfnamefont {C.}~\bibnamefont {Sayrin}},
  \bibinfo {author} {\bibfnamefont {B.}~\bibnamefont {Albrecht}}, \bibinfo
  {author} {\bibfnamefont {P.}~\bibnamefont {Schneeweiss}}, \ and\ \bibinfo
  {author} {\bibfnamefont {A.}~\bibnamefont {Rauschenbeutel}},\ }\href
  {\doibase 10.1038/ncomms6713} {\bibfield  {journal} {\bibinfo  {journal}
  {Nat. Commun.}\ }\textbf {\bibinfo {volume} {5}},\ \bibinfo {pages} {5713}
  (\bibinfo {year} {2014})}\BibitemShut {NoStop}%
\bibitem [{\citenamefont {Petersen}\ \emph {et~al.}(2014)\citenamefont
  {Petersen}, \citenamefont {Volz},\ and\ \citenamefont
  {Rauschenbeutel}}]{Petersen14}%
  \BibitemOpen
  \bibfield  {author} {\bibinfo {author} {\bibfnamefont {J.}~\bibnamefont
  {Petersen}}, \bibinfo {author} {\bibfnamefont {J.}~\bibnamefont {Volz}}, \
  and\ \bibinfo {author} {\bibfnamefont {A.}~\bibnamefont {Rauschenbeutel}},\
  }\href@noop {} {\bibfield  {journal} {\bibinfo  {journal} {Science}\ }\textbf
  {\bibinfo {volume} {346}},\ \bibinfo {pages} {67} (\bibinfo {year}
  {2014})}\BibitemShut {NoStop}%
\bibitem [{\citenamefont {le~Feber}\ \emph {et~al.}(2015)\citenamefont
  {le~Feber}, \citenamefont {Rotenberg},\ and\ \citenamefont
  {Kuipers}}]{Feber15}%
  \BibitemOpen
  \bibfield  {author} {\bibinfo {author} {\bibfnamefont {B.}~\bibnamefont
  {le~Feber}}, \bibinfo {author} {\bibfnamefont {N.}~\bibnamefont {Rotenberg}},
  \ and\ \bibinfo {author} {\bibfnamefont {L.}~\bibnamefont {Kuipers}},\ }\href
  {\doibase 10.1038/ncomms7695} {\bibfield  {journal} {\bibinfo  {journal}
  {Nat. Commun.}\ }\textbf {\bibinfo {volume} {6}},\ \bibinfo {pages} {6695}
  (\bibinfo {year} {2015})}\BibitemShut {NoStop}%
\bibitem [{\citenamefont {{S{\"o}llner}}\ \emph {et~al.}(2014)\citenamefont
  {{S{\"o}llner}}, \citenamefont {{Mahmoodian}}, \citenamefont {{Lindskov
  Hansen}}, \citenamefont {{Midolo}}, \citenamefont {{Javadi}}, \citenamefont
  {{Kir{\v s}ansk{\.e}}}, \citenamefont {{Pregnolato}}, \citenamefont
  {{El-Ella}}, \citenamefont {{Hye Lee}}, \citenamefont {{Song}}, \citenamefont
  {{Stobbe}},\ and\ \citenamefont {{Lodahl}}}]{ArXiv_Soellner14}%
  \BibitemOpen
  \bibfield  {author} {\bibinfo {author} {\bibfnamefont {I.}~\bibnamefont
  {{S{\"o}llner}}}, \bibinfo {author} {\bibfnamefont {S.}~\bibnamefont
  {{Mahmoodian}}}, \bibinfo {author} {\bibfnamefont {S.}~\bibnamefont
  {{Lindskov Hansen}}}, \bibinfo {author} {\bibfnamefont {L.}~\bibnamefont
  {{Midolo}}}, \bibinfo {author} {\bibfnamefont {A.}~\bibnamefont {{Javadi}}},
  \bibinfo {author} {\bibfnamefont {G.}~\bibnamefont {{Kir{\v s}ansk{\.e}}}},
  \bibinfo {author} {\bibfnamefont {T.}~\bibnamefont {{Pregnolato}}}, \bibinfo
  {author} {\bibfnamefont {H.}~\bibnamefont {{El-Ella}}}, \bibinfo {author}
  {\bibfnamefont {E.}~\bibnamefont {{Hye Lee}}}, \bibinfo {author}
  {\bibfnamefont {J.~D.}\ \bibnamefont {{Song}}}, \bibinfo {author}
  {\bibfnamefont {S.}~\bibnamefont {{Stobbe}}}, \ and\ \bibinfo {author}
  {\bibfnamefont {P.}~\bibnamefont {{Lodahl}}},\ }\href@noop {} {\  (\bibinfo
  {year} {2014})},\ \Eprint {http://arxiv.org/abs/1406.4295} {arXiv:1406.4295}
  \BibitemShut {NoStop}%
\bibitem [{\citenamefont {Xi}\ \emph {et~al.}(2013)\citenamefont {Xi},
  \citenamefont {Lu}, \citenamefont {Yao}, \citenamefont {Yu}, \citenamefont
  {Wang},\ and\ \citenamefont {Ming}}]{Xi13}%
  \BibitemOpen
  \bibfield  {author} {\bibinfo {author} {\bibfnamefont {Z.}~\bibnamefont
  {Xi}}, \bibinfo {author} {\bibfnamefont {Y.}~\bibnamefont {Lu}}, \bibinfo
  {author} {\bibfnamefont {P.}~\bibnamefont {Yao}}, \bibinfo {author}
  {\bibfnamefont {W.}~\bibnamefont {Yu}}, \bibinfo {author} {\bibfnamefont
  {P.}~\bibnamefont {Wang}}, \ and\ \bibinfo {author} {\bibfnamefont
  {H.}~\bibnamefont {Ming}},\ }\href {\doibase 10.1364/OE.21.030327} {\bibfield
   {journal} {\bibinfo  {journal} {Opt. Express}\ }\textbf {\bibinfo {volume}
  {21}},\ \bibinfo {pages} {30327} (\bibinfo {year} {2013})}\BibitemShut
  {NoStop}%
\bibitem [{\citenamefont {Dalvit}\ \emph {et~al.}(2008)\citenamefont {Dalvit},
  \citenamefont {Neto}, \citenamefont {Lambrecht},\ and\ \citenamefont
  {Reynaud}}]{Dalvit08}%
  \BibitemOpen
  \bibfield  {author} {\bibinfo {author} {\bibfnamefont {D.~A.~R.}\
  \bibnamefont {Dalvit}}, \bibinfo {author} {\bibfnamefont {P.~A.~M.}\
  \bibnamefont {Neto}}, \bibinfo {author} {\bibfnamefont {A.}~\bibnamefont
  {Lambrecht}}, \ and\ \bibinfo {author} {\bibfnamefont {S.}~\bibnamefont
  {Reynaud}},\ }\href {\doibase 10.1103/PhysRevLett.100.040405} {\bibfield
  {journal} {\bibinfo  {journal} {Phys. Rev. Lett.}\ }\textbf {\bibinfo
  {volume} {100}},\ \bibinfo {pages} {040405} (\bibinfo {year}
  {2008})}\BibitemShut {NoStop}%
\bibitem [{\citenamefont {D\"obrich}\ \emph {et~al.}(2008)\citenamefont
  {D\"obrich}, \citenamefont {DeKieviet},\ and\ \citenamefont
  {Gies}}]{Doebrich08}%
  \BibitemOpen
  \bibfield  {author} {\bibinfo {author} {\bibfnamefont {B.}~\bibnamefont
  {D\"obrich}}, \bibinfo {author} {\bibfnamefont {M.}~\bibnamefont
  {DeKieviet}}, \ and\ \bibinfo {author} {\bibfnamefont {H.}~\bibnamefont
  {Gies}},\ }\href {\doibase 10.1103/PhysRevD.78.125022} {\bibfield  {journal}
  {\bibinfo  {journal} {Phys. Rev. D}\ }\textbf {\bibinfo {volume} {78}},\
  \bibinfo {pages} {125022} (\bibinfo {year} {2008})}\BibitemShut {NoStop}%
\bibitem [{\citenamefont {Messina}\ \emph {et~al.}(2009)\citenamefont
  {Messina}, \citenamefont {Dalvit}, \citenamefont {Neto}, \citenamefont
  {Lambrecht},\ and\ \citenamefont {Reynaud}}]{Messina09}%
  \BibitemOpen
  \bibfield  {author} {\bibinfo {author} {\bibfnamefont {R.}~\bibnamefont
  {Messina}}, \bibinfo {author} {\bibfnamefont {D.~A.~R.}\ \bibnamefont
  {Dalvit}}, \bibinfo {author} {\bibfnamefont {P.~A.~M.}\ \bibnamefont {Neto}},
  \bibinfo {author} {\bibfnamefont {A.}~\bibnamefont {Lambrecht}}, \ and\
  \bibinfo {author} {\bibfnamefont {S.}~\bibnamefont {Reynaud}},\ }\href
  {\doibase 10.1103/PhysRevA.80.022119} {\bibfield  {journal} {\bibinfo
  {journal} {Phys. Rev. A}\ }\textbf {\bibinfo {volume} {80}},\ \bibinfo
  {pages} {022119} (\bibinfo {year} {2009})}\BibitemShut {NoStop}%
\bibitem [{\citenamefont {Contreras-Reyes}\ \emph {et~al.}(2010)\citenamefont
  {Contreras-Reyes}, \citenamefont {Gu\'erout}, \citenamefont {Neto},
  \citenamefont {Dalvit}, \citenamefont {Lambrecht},\ and\ \citenamefont
  {Reynaud}}]{Contreras10}%
  \BibitemOpen
  \bibfield  {author} {\bibinfo {author} {\bibfnamefont {A.~M.}\ \bibnamefont
  {Contreras-Reyes}}, \bibinfo {author} {\bibfnamefont {R.}~\bibnamefont
  {Gu\'erout}}, \bibinfo {author} {\bibfnamefont {P.~A.~M.}\ \bibnamefont
  {Neto}}, \bibinfo {author} {\bibfnamefont {D.~A.~R.}\ \bibnamefont {Dalvit}},
  \bibinfo {author} {\bibfnamefont {A.}~\bibnamefont {Lambrecht}}, \ and\
  \bibinfo {author} {\bibfnamefont {S.}~\bibnamefont {Reynaud}},\ }\href
  {\doibase 10.1103/PhysRevA.82.052517} {\bibfield  {journal} {\bibinfo
  {journal} {Phys. Rev. A}\ }\textbf {\bibinfo {volume} {82}},\ \bibinfo
  {pages} {052517} (\bibinfo {year} {2010})}\BibitemShut {NoStop}%
\bibitem [{\citenamefont {Moreno}\ \emph {et~al.}(2010)\citenamefont {Moreno},
  \citenamefont {Messina}, \citenamefont {Dalvit}, \citenamefont {Lambrecht},
  \citenamefont {Neto},\ and\ \citenamefont {Reynaud}}]{Moreno10}%
  \BibitemOpen
  \bibfield  {author} {\bibinfo {author} {\bibfnamefont {G.~A.}\ \bibnamefont
  {Moreno}}, \bibinfo {author} {\bibfnamefont {R.}~\bibnamefont {Messina}},
  \bibinfo {author} {\bibfnamefont {D.~A.~R.}\ \bibnamefont {Dalvit}}, \bibinfo
  {author} {\bibfnamefont {A.}~\bibnamefont {Lambrecht}}, \bibinfo {author}
  {\bibfnamefont {P.~A.~M.}\ \bibnamefont {Neto}}, \ and\ \bibinfo {author}
  {\bibfnamefont {S.}~\bibnamefont {Reynaud}},\ }\href {\doibase
  10.1103/PhysRevLett.105.210401} {\bibfield  {journal} {\bibinfo  {journal}
  {Phys. Rev. Lett.}\ }\textbf {\bibinfo {volume} {105}},\ \bibinfo {pages}
  {210401} (\bibinfo {year} {2010})}\BibitemShut {NoStop}%
\bibitem [{\citenamefont {Rodrigues}\ \emph {et~al.}(2006)\citenamefont
  {Rodrigues}, \citenamefont {Neto}, \citenamefont {Lambrecht},\ and\
  \citenamefont {Reynaud}}]{Rodrigues06}%
  \BibitemOpen
  \bibfield  {author} {\bibinfo {author} {\bibfnamefont {R.~B.}\ \bibnamefont
  {Rodrigues}}, \bibinfo {author} {\bibfnamefont {P.~A.~M.}\ \bibnamefont
  {Neto}}, \bibinfo {author} {\bibfnamefont {A.}~\bibnamefont {Lambrecht}}, \
  and\ \bibinfo {author} {\bibfnamefont {S.}~\bibnamefont {Reynaud}},\ }\href
  {\doibase 10.1103/PhysRevLett.96.100402} {\bibfield  {journal} {\bibinfo
  {journal} {Phys. Rev. Lett.}\ }\textbf {\bibinfo {volume} {96}},\ \bibinfo
  {pages} {100402} (\bibinfo {year} {2006})}\BibitemShut {NoStop}%
\bibitem [{\citenamefont {Lambrecht}\ and\ \citenamefont
  {Marachevsky}(2008)}]{Lambrecht08}%
  \BibitemOpen
  \bibfield  {author} {\bibinfo {author} {\bibfnamefont {A.}~\bibnamefont
  {Lambrecht}}\ and\ \bibinfo {author} {\bibfnamefont {V.~N.}\ \bibnamefont
  {Marachevsky}},\ }\href {\doibase 10.1103/PhysRevLett.101.160403} {\bibfield
  {journal} {\bibinfo  {journal} {Phys. Rev. Lett.}\ }\textbf {\bibinfo
  {volume} {101}},\ \bibinfo {pages} {160403} (\bibinfo {year}
  {2008})}\BibitemShut {NoStop}%
\bibitem [{\citenamefont {Chiu}\ \emph {et~al.}(2009)\citenamefont {Chiu},
  \citenamefont {Klimchitskaya}, \citenamefont {Marachevsky}, \citenamefont
  {Mostepanenko},\ and\ \citenamefont {Mohideen}}]{Chiu09}%
  \BibitemOpen
  \bibfield  {author} {\bibinfo {author} {\bibfnamefont {H.-C.}\ \bibnamefont
  {Chiu}}, \bibinfo {author} {\bibfnamefont {G.~L.}\ \bibnamefont
  {Klimchitskaya}}, \bibinfo {author} {\bibfnamefont {V.~N.}\ \bibnamefont
  {Marachevsky}}, \bibinfo {author} {\bibfnamefont {V.~M.}\ \bibnamefont
  {Mostepanenko}}, \ and\ \bibinfo {author} {\bibfnamefont {U.}~\bibnamefont
  {Mohideen}},\ }\href {\doibase 10.1103/PhysRevB.80.121402} {\bibfield
  {journal} {\bibinfo  {journal} {Phys. Rev. B}\ }\textbf {\bibinfo {volume}
  {80}},\ \bibinfo {pages} {121402} (\bibinfo {year} {2009})}\BibitemShut
  {NoStop}%
\bibitem [{\citenamefont {Chen}\ \emph {et~al.}(2002)\citenamefont {Chen},
  \citenamefont {Mohideen}, \citenamefont {Klimchitskaya},\ and\ \citenamefont
  {Mostepanenko}}]{Chen02}%
  \BibitemOpen
  \bibfield  {author} {\bibinfo {author} {\bibfnamefont {F.}~\bibnamefont
  {Chen}}, \bibinfo {author} {\bibfnamefont {U.}~\bibnamefont {Mohideen}},
  \bibinfo {author} {\bibfnamefont {G.~L.}\ \bibnamefont {Klimchitskaya}}, \
  and\ \bibinfo {author} {\bibfnamefont {V.~M.}\ \bibnamefont {Mostepanenko}},\
  }\href {\doibase 10.1103/PhysRevLett.88.101801} {\bibfield  {journal}
  {\bibinfo  {journal} {Phys. Rev. Lett.}\ }\textbf {\bibinfo {volume} {88}},\
  \bibinfo {pages} {101801} (\bibinfo {year} {2002})}\BibitemShut {NoStop}%
\bibitem [{\citenamefont {Ashourvan}\ \emph {et~al.}(2007)\citenamefont
  {Ashourvan}, \citenamefont {Miri},\ and\ \citenamefont
  {Golestanian}}]{Ashourvan07}%
  \BibitemOpen
  \bibfield  {author} {\bibinfo {author} {\bibfnamefont {A.}~\bibnamefont
  {Ashourvan}}, \bibinfo {author} {\bibfnamefont {M.~F.}\ \bibnamefont {Miri}},
  \ and\ \bibinfo {author} {\bibfnamefont {R.}~\bibnamefont {Golestanian}},\
  }\href {\doibase 10.1103/PhysRevLett.98.140801} {\bibfield  {journal}
  {\bibinfo  {journal} {Phys. Rev. Lett.}\ }\textbf {\bibinfo {volume} {98}},\
  \bibinfo {pages} {140801} (\bibinfo {year} {2007})}\BibitemShut {NoStop}%
\bibitem [{\citenamefont {Steck}(2010)}]{Steck10}%
  \BibitemOpen
  \bibfield  {author} {\bibinfo {author} {\bibfnamefont {D.~A.}\ \bibnamefont
  {Steck}},\ }\href@noop {} {\bibfield  {journal} {\bibinfo  {journal}
  {available online at http://steck.us/alkalidata}\ } (\bibinfo {year}
  {2010})}\BibitemShut {NoStop}%
\bibitem [{\citenamefont {{Fam Le Kien}}\ \emph {et~al.}(2005)\citenamefont
  {{Fam Le Kien}}, \citenamefont {Dutta~Gupta}, \citenamefont {Balykin},\ and\
  \citenamefont {Hakuta}}]{LeKien05c}%
  \BibitemOpen
  \bibfield  {author} {\bibinfo {author} {\bibnamefont {{Fam Le Kien}}},
  \bibinfo {author} {\bibfnamefont {S.}~\bibnamefont {Dutta~Gupta}}, \bibinfo
  {author} {\bibfnamefont {V.~I.}\ \bibnamefont {Balykin}}, \ and\ \bibinfo
  {author} {\bibfnamefont {K.}~\bibnamefont {Hakuta}},\ }\href {\doibase
  10.1103/PhysRevA.72.032509} {\bibfield  {journal} {\bibinfo  {journal} {Phys.
  Rev. A}\ }\textbf {\bibinfo {volume} {72}},\ \bibinfo {pages} {032509}
  (\bibinfo {year} {2005})}\BibitemShut {NoStop}%
\bibitem [{\citenamefont {{See Supplemental Material at [URL will be inserted
  by publisher] for derivations}}()}]{Supplement}%
  \BibitemOpen
  \bibfield  {author} {\bibinfo {author} {\bibnamefont {{See Supplemental
  Material at [URL will be inserted by publisher] for derivations}}},\
  }\href@noop {} {\ }\BibitemShut {NoStop}%
\bibitem [{\citenamefont {Le~Kien}\ and\ \citenamefont
  {Rauschenbeutel}(2014)}]{LeKien14a}%
  \BibitemOpen
  \bibfield  {author} {\bibinfo {author} {\bibfnamefont {F.}~\bibnamefont
  {Le~Kien}}\ and\ \bibinfo {author} {\bibfnamefont {A.}~\bibnamefont
  {Rauschenbeutel}},\ }\href {\doibase 10.1103/PhysRevA.90.023805} {\bibfield
  {journal} {\bibinfo  {journal} {Phys. Rev. A}\ }\textbf {\bibinfo {volume}
  {90}},\ \bibinfo {pages} {023805} (\bibinfo {year} {2014})}\BibitemShut
  {NoStop}%
\bibitem [{\citenamefont {Drexhage}(1970)}]{Drexhage70}%
  \BibitemOpen
  \bibfield  {author} {\bibinfo {author} {\bibfnamefont {K.}~\bibnamefont
  {Drexhage}},\ }\href@noop {} {\bibfield  {journal} {\bibinfo  {journal} {J.
  Lumin.}\ }\textbf {\bibinfo {volume} {1}},\ \bibinfo {pages} {693} (\bibinfo
  {year} {1970})}\BibitemShut {NoStop}%
\bibitem [{\citenamefont {Buhmann}\ \emph {et~al.}(2004)\citenamefont
  {Buhmann}, \citenamefont {Kn\"{o}ll}, \citenamefont {Welsch},\ and\
  \citenamefont {Dung}}]{Buhmann04}%
  \BibitemOpen
  \bibfield  {author} {\bibinfo {author} {\bibfnamefont {S.~Y.}\ \bibnamefont
  {Buhmann}}, \bibinfo {author} {\bibfnamefont {L.}~\bibnamefont {Kn\"{o}ll}},
  \bibinfo {author} {\bibfnamefont {D.-G.}\ \bibnamefont {Welsch}}, \ and\
  \bibinfo {author} {\bibfnamefont {H.~T.}\ \bibnamefont {Dung}},\ }\href@noop
  {} {\bibfield  {journal} {\bibinfo  {journal} {Phys. Rev. A}\ }\textbf
  {\bibinfo {volume} {70}},\ \bibinfo {pages} {052117} (\bibinfo {year}
  {2004})}\BibitemShut {NoStop}%
\bibitem [{\citenamefont {Li}\ \emph {et~al.}(2000)\citenamefont {Li},
  \citenamefont {Leong}, \citenamefont {Yeo},\ and\ \citenamefont
  {Kooi}}]{Li00}%
  \BibitemOpen
  \bibfield  {author} {\bibinfo {author} {\bibfnamefont {L.-W.}\ \bibnamefont
  {Li}}, \bibinfo {author} {\bibfnamefont {M.-S.}\ \bibnamefont {Leong}},
  \bibinfo {author} {\bibfnamefont {T.-S.}\ \bibnamefont {Yeo}}, \ and\
  \bibinfo {author} {\bibfnamefont {P.-S.}\ \bibnamefont {Kooi}},\ }\href@noop
  {} {\bibfield  {journal} {\bibinfo  {journal} {J. Electromagn. Wave.}\
  }\textbf {\bibinfo {volume} {14}},\ \bibinfo {pages} {961} (\bibinfo {year}
  {2000})}\BibitemShut {NoStop}%
\bibitem [{\citenamefont {Kitamura}\ \emph {et~al.}(2007)\citenamefont
  {Kitamura}, \citenamefont {Pilon},\ and\ \citenamefont
  {Jonasz}}]{Kitamura07}%
  \BibitemOpen
  \bibfield  {author} {\bibinfo {author} {\bibfnamefont {R.}~\bibnamefont
  {Kitamura}}, \bibinfo {author} {\bibfnamefont {L.}~\bibnamefont {Pilon}}, \
  and\ \bibinfo {author} {\bibfnamefont {M.}~\bibnamefont {Jonasz}},\ }\href
  {\doibase 10.1364/AO.46.008118} {\bibfield  {journal} {\bibinfo  {journal}
  {Appl. Opt.}\ }\textbf {\bibinfo {volume} {46}},\ \bibinfo {pages} {8118}
  (\bibinfo {year} {2007})}\BibitemShut {NoStop}%
\bibitem [{\citenamefont {Neuman}\ and\ \citenamefont
  {Block}(2004)}]{Neuman04}%
  \BibitemOpen
  \bibfield  {author} {\bibinfo {author} {\bibfnamefont {K.~C.}\ \bibnamefont
  {Neuman}}\ and\ \bibinfo {author} {\bibfnamefont {S.~M.}\ \bibnamefont
  {Block}},\ }\href@noop {} {\bibfield  {journal} {\bibinfo  {journal} {Rev.
  Sci. Instrum.}\ }\textbf {\bibinfo {volume} {75}},\ \bibinfo {pages} {2787}
  (\bibinfo {year} {2004})}\BibitemShut {NoStop}%
\bibitem [{\citenamefont {Chang}\ \emph {et~al.}(2013)\citenamefont {Chang},
  \citenamefont {Cirac},\ and\ \citenamefont {Kimble}}]{Chang13}%
  \BibitemOpen
  \bibfield  {author} {\bibinfo {author} {\bibfnamefont {D.~E.}\ \bibnamefont
  {Chang}}, \bibinfo {author} {\bibfnamefont {J.~I.}\ \bibnamefont {Cirac}}, \
  and\ \bibinfo {author} {\bibfnamefont {H.~J.}\ \bibnamefont {Kimble}},\
  }\href {\doibase 10.1103/PhysRevLett.110.113606} {\bibfield  {journal}
  {\bibinfo  {journal} {Phys. Rev. Lett.}\ }\textbf {\bibinfo {volume} {110}},\
  \bibinfo {pages} {113606} (\bibinfo {year} {2013})}\BibitemShut {NoStop}%
\bibitem [{\citenamefont {Holzmann}\ \emph {et~al.}(2014)\citenamefont
  {Holzmann}, \citenamefont {Sonnleitner},\ and\ \citenamefont
  {Ritsch}}]{Holzmann14}%
  \BibitemOpen
  \bibfield  {author} {\bibinfo {author} {\bibfnamefont {D.}~\bibnamefont
  {Holzmann}}, \bibinfo {author} {\bibfnamefont {M.}~\bibnamefont
  {Sonnleitner}}, \ and\ \bibinfo {author} {\bibfnamefont {H.}~\bibnamefont
  {Ritsch}},\ }\href@noop {} {\bibfield  {journal} {\bibinfo  {journal} {Eur.
  Phys. J. D}\ }\textbf {\bibinfo {volume} {68}},\ \bibinfo {pages} {1}
  (\bibinfo {year} {2014})}\BibitemShut {NoStop}%
\bibitem [{\citenamefont {Junge}\ \emph {et~al.}(2013)\citenamefont {Junge},
  \citenamefont {O'Shea}, \citenamefont {Volz},\ and\ \citenamefont
  {Rauschenbeutel}}]{Junge13}%
  \BibitemOpen
  \bibfield  {author} {\bibinfo {author} {\bibfnamefont {C.}~\bibnamefont
  {Junge}}, \bibinfo {author} {\bibfnamefont {D.}~\bibnamefont {O'Shea}},
  \bibinfo {author} {\bibfnamefont {J.}~\bibnamefont {Volz}}, \ and\ \bibinfo
  {author} {\bibfnamefont {A.}~\bibnamefont {Rauschenbeutel}},\ }\href
  {\doibase 10.1103/PhysRevLett.110.213604} {\bibfield  {journal} {\bibinfo
  {journal} {Phys. Rev. Lett.}\ }\textbf {\bibinfo {volume} {110}},\ \bibinfo
  {pages} {213604} (\bibinfo {year} {2013})}\BibitemShut {NoStop}%
\bibitem [{\citenamefont {Shomroni}\ \emph {et~al.}(2014)\citenamefont
  {Shomroni}, \citenamefont {Rosenblum}, \citenamefont {Lovsky}, \citenamefont
  {Bechler}, \citenamefont {Guendelman},\ and\ \citenamefont
  {Dayan}}]{Shomroni14}%
  \BibitemOpen
  \bibfield  {author} {\bibinfo {author} {\bibfnamefont {I.}~\bibnamefont
  {Shomroni}}, \bibinfo {author} {\bibfnamefont {S.}~\bibnamefont {Rosenblum}},
  \bibinfo {author} {\bibfnamefont {Y.}~\bibnamefont {Lovsky}}, \bibinfo
  {author} {\bibfnamefont {O.}~\bibnamefont {Bechler}}, \bibinfo {author}
  {\bibfnamefont {G.}~\bibnamefont {Guendelman}}, \ and\ \bibinfo {author}
  {\bibfnamefont {B.}~\bibnamefont {Dayan}},\ }\href@noop {} {\bibfield
  {journal} {\bibinfo  {journal} {Science}\ }\textbf {\bibinfo {volume}
  {345}},\ \bibinfo {pages} {903} (\bibinfo {year} {2014})}\BibitemShut
  {NoStop}%
\bibitem [{\citenamefont {Hammes}\ \emph {et~al.}(2003)\citenamefont {Hammes},
  \citenamefont {Rychtarik}, \citenamefont {Engeser}, \citenamefont
  {N\"agerl},\ and\ \citenamefont {Grimm}}]{Hammes03}%
  \BibitemOpen
  \bibfield  {author} {\bibinfo {author} {\bibfnamefont {M.}~\bibnamefont
  {Hammes}}, \bibinfo {author} {\bibfnamefont {D.}~\bibnamefont {Rychtarik}},
  \bibinfo {author} {\bibfnamefont {B.}~\bibnamefont {Engeser}}, \bibinfo
  {author} {\bibfnamefont {H.-C.}\ \bibnamefont {N\"agerl}}, \ and\ \bibinfo
  {author} {\bibfnamefont {R.}~\bibnamefont {Grimm}},\ }\href {\doibase
  10.1103/PhysRevLett.90.173001} {\bibfield  {journal} {\bibinfo  {journal}
  {Phys. Rev. Lett.}\ }\textbf {\bibinfo {volume} {90}},\ \bibinfo {pages}
  {173001} (\bibinfo {year} {2003})}\BibitemShut {NoStop}%
\bibitem [{\citenamefont {{Stehle}}\ \emph {et~al.}(2014)\citenamefont
  {{Stehle}}, \citenamefont {{Zimmermann}},\ and\ \citenamefont
  {{Slama}}}]{Stehle14}%
  \BibitemOpen
  \bibfield  {author} {\bibinfo {author} {\bibfnamefont {C.}~\bibnamefont
  {{Stehle}}}, \bibinfo {author} {\bibfnamefont {C.}~\bibnamefont
  {{Zimmermann}}}, \ and\ \bibinfo {author} {\bibfnamefont {S.}~\bibnamefont
  {{Slama}}},\ }\href {\doibase 10.1038/nphys3129} {\bibfield  {journal}
  {\bibinfo  {journal} {Nat. Phys.}\ }\textbf {\bibinfo {volume} {10}},\
  \bibinfo {pages} {937} (\bibinfo {year} {2014})}\BibitemShut {NoStop}%
\bibitem [{\citenamefont {Vetsch}\ \emph {et~al.}(2012)\citenamefont {Vetsch},
  \citenamefont {Dawkins}, \citenamefont {Mitsch}, \citenamefont {Reitz},
  \citenamefont {Schneeweiss},\ and\ \citenamefont
  {Rauschenbeutel}}]{Vetsch12}%
  \BibitemOpen
  \bibfield  {author} {\bibinfo {author} {\bibfnamefont {E.}~\bibnamefont
  {Vetsch}}, \bibinfo {author} {\bibfnamefont {S.}~\bibnamefont {Dawkins}},
  \bibinfo {author} {\bibfnamefont {R.}~\bibnamefont {Mitsch}}, \bibinfo
  {author} {\bibfnamefont {D.}~\bibnamefont {Reitz}}, \bibinfo {author}
  {\bibfnamefont {P.}~\bibnamefont {Schneeweiss}}, \ and\ \bibinfo {author}
  {\bibfnamefont {A.}~\bibnamefont {Rauschenbeutel}},\ }\href {\doibase
  10.1109/JSTQE.2012.2196025} {\bibfield  {journal} {\bibinfo  {journal} {IEEE
  J. Quantum Electron.}\ }\textbf {\bibinfo {volume} {18}},\ \bibinfo {pages}
  {1763 } (\bibinfo {year} {2012})}\BibitemShut {NoStop}%
\bibitem [{\citenamefont {Thompson}\ \emph {et~al.}(2013)\citenamefont
  {Thompson}, \citenamefont {Tiecke}, \citenamefont {de~Leon}, \citenamefont
  {Feist}, \citenamefont {Akimov}, \citenamefont {Gullans}, \citenamefont
  {Zibrov}, \citenamefont {Vuletic},\ and\ \citenamefont
  {Lukin}}]{Thompson13b}%
  \BibitemOpen
  \bibfield  {author} {\bibinfo {author} {\bibfnamefont {J.~D.}\ \bibnamefont
  {Thompson}}, \bibinfo {author} {\bibfnamefont {T.~G.}\ \bibnamefont
  {Tiecke}}, \bibinfo {author} {\bibfnamefont {N.~P.}\ \bibnamefont {de~Leon}},
  \bibinfo {author} {\bibfnamefont {J.}~\bibnamefont {Feist}}, \bibinfo
  {author} {\bibfnamefont {A.~V.}\ \bibnamefont {Akimov}}, \bibinfo {author}
  {\bibfnamefont {M.}~\bibnamefont {Gullans}}, \bibinfo {author} {\bibfnamefont
  {A.~S.}\ \bibnamefont {Zibrov}}, \bibinfo {author} {\bibfnamefont
  {V.}~\bibnamefont {Vuletic}}, \ and\ \bibinfo {author} {\bibfnamefont
  {M.~D.}\ \bibnamefont {Lukin}},\ }\href {\doibase 10.1126/science.1237125}
  {\bibfield  {journal} {\bibinfo  {journal} {Science}\ }\textbf {\bibinfo
  {volume} {340}},\ \bibinfo {pages} {1202} (\bibinfo {year}
  {2013})}\BibitemShut {NoStop}%
\bibitem [{\citenamefont {Goban}\ \emph {et~al.}(2014)\citenamefont {Goban},
  \citenamefont {Hung}, \citenamefont {Yu}, \citenamefont {Hood}, \citenamefont
  {Muniz}, \citenamefont {Lee}, \citenamefont {Martin}, \citenamefont
  {McClung}, \citenamefont {Choi}, \citenamefont {Chang}, \citenamefont
  {Painter},\ and\ \citenamefont {Kimble}}]{Goban14}%
  \BibitemOpen
  \bibfield  {author} {\bibinfo {author} {\bibfnamefont {A.}~\bibnamefont
  {Goban}}, \bibinfo {author} {\bibfnamefont {C.-L.}\ \bibnamefont {Hung}},
  \bibinfo {author} {\bibfnamefont {S.-P.}\ \bibnamefont {Yu}}, \bibinfo
  {author} {\bibfnamefont {J.}~\bibnamefont {Hood}}, \bibinfo {author}
  {\bibfnamefont {J.}~\bibnamefont {Muniz}}, \bibinfo {author} {\bibfnamefont
  {J.}~\bibnamefont {Lee}}, \bibinfo {author} {\bibfnamefont {M.}~\bibnamefont
  {Martin}}, \bibinfo {author} {\bibfnamefont {A.}~\bibnamefont {McClung}},
  \bibinfo {author} {\bibfnamefont {K.}~\bibnamefont {Choi}}, \bibinfo {author}
  {\bibfnamefont {D.}~\bibnamefont {Chang}}, \bibinfo {author} {\bibfnamefont
  {O.}~\bibnamefont {Painter}}, \ and\ \bibinfo {author} {\bibfnamefont
  {H.}~\bibnamefont {Kimble}},\ }\href {http://dx.doi.org/10.1038/ncomms4808}
  {\bibfield  {journal} {\bibinfo  {journal} {Nat. Commun.}\ }\textbf {\bibinfo
  {volume} {5}},\  (\bibinfo {year} {2014})}\BibitemShut {NoStop}%
\end{thebibliography}%

\end{document}